\documentclass[12pt]{article}
\usepackage{amssymb,amsmath,slashed,latexsym}
\newcommand{\ft}[2]{{\textstyle\frac{#1}{#2}}}

\def\tr{\mathop{\rm tr}\nolimits}
\def\rme{{\rm e}}

\newcommand{\hc}{{\rm h.c.}}
\newsavebox{\uuunit}
\sbox{\uuunit}
    {\setlength{\unitlength}{0.825em}
     \begin{picture}(0.6,0.7)
        \thinlines
        \put(0,0){\line(1,0){0.5}}
        \put(0.15,0){\line(0,1){0.7}}
        \put(0.35,0){\line(0,1){0.8}}
       \multiput(0.3,0.8)(-0.04,-0.02){12}{\rule{0.5pt}{0.5pt}}
     \end {picture}}

\csname @addtoreset\endcsname{equation}{section}
   \usepackage[nosort]{cite}
   \pdfoutput=1
  \usepackage[pdftex]{hyperref}
\usepackage{color}
\begin{document}

 %%%%%%%%%%%%%%%%%%%%%%%%%%%%%%%%%%%%%%%%%%%%%%%%%%%%%%%%%%%
\begin{titlepage}
\begin{flushright}
CERN-TH-2016-204
\end{flushright}
%\phantom{.}
\vspace{.5cm}
\begin{center}
\baselineskip=16pt
{\LARGE    Mass Formulae for Broken Supersymmetry \\ \vskip 0.2cm in Curved Space-Time % \\ \vskip 0.2cm between lines
}\\
\vfill%\vskip 15mm%27.mm
{\large  {\bf Sergio Ferrara}$^{1,2,3}$, and {\large  \bf Antoine Van Proeyen$^4$}, } \\
\vfill

{\small$^1$ Theoretical Physics Department, CERN CH-1211 Geneva 23, Switzerland\\\smallskip
$^2$ INFN - Laboratori Nazionali di Frascati Via Enrico Fermi 40, I-00044 Frascati, Italy\\\smallskip
$^3$ Department of Physics and Astronomy, U.C.L.A., Los Angeles CA 90095-1547, USA\\\smallskip
$^4$   KU Leuven, Institute for Theoretical Physics, Celestijnenlaan 200D, B-3001 Leuven,
Belgium  \\[2mm] }
\end{center}
\vfill
\begin{center}
{\bf Abstract}
\end{center}
{\small
We derive the mass formulae for ${\cal N}=1$, $D=4$ matter-coupled Supergravity for broken (and
unbroken) Supersymmetry in curved space-time. These formulae are applicable to
De Sitter configurations as is the case for inflation. For unbroken Supersymmetry
in anti-de Sitter (AdS) one gets the mass relations modified by the AdS curvature.
We compute the mass relations both for the potential and its derivative non-vanishing.
}\vspace{2mm} \vfill \hrule width 3.cm
{\footnotesize \noindent e-mails: Sergio.Ferrara@cern.ch,
antoine.vanproeyen@fys.kuleuven.be }\\
Paper to be published in Fortschritte der Physik.
\end{titlepage}
\addtocounter{page}{1}
 \tableofcontents{}
\newpage
%%%%%%%%%%%%%%%%
\section{Introduction}
In the 80's the matter-coupled ${\cal N}=1$ supergravity theories were constructed. At that time, the physics community was considering
a Minkowski spacetime. One of the main results of \cite{Ferrara:1979wa,Cremmer:1982en,Grisaru:1982sr} was the supertrace formulae.  These mass formulae were important for phenomenological model building. The result was given for a Minkowski background. Meanwhile we know that spacetime can be de Sitter, and thus a generalization of the result of 1982 is required.

We consider mass formulae for theories with $N$ chiral multiplets coupled to supergravity for arbitrary background potentials and for non-extremal points.

The formulation is given in terms of the K\"{a}hler-invariant function ${\cal G}(z,\bar z)$ and in the formulation with K\"{a}hler potential ${\cal K}(z,\bar z)$ and superpotential $W(z)$, where $z$ represent the complex scalar fields. The relation between the two formalisms is given in Appendix \ref{app:GKW}, where also some basic quantities are defined. We further use all notations from \cite{Freedman:2012zz}.

In Sec. \ref{ss:superBEH} we consider the definition of the Goldstino in curved space leading to the super-Brout-Englert-Higgs (BEH) formalism. This defines the effective mass matrix for the physical fermions. The mass matrices are calculated in Sec.~\ref{ss:Str}, where the main result is obtained.
Concluding remarks are given in Sec. \ref{ss:conclusion}.

\section{Super-BEH formalism in curved space}
\label{ss:superBEH}
The super-BEH effect in Minkowski space was considered in \cite{Cremmer:1982en,Grisaru:1982sr}. It was generalised to a cosmological background (with non-constant scalars)  in \cite{Kallosh:2000ve}. We will now give the steps for a more general situation. The notations are given in Appendix~\ref{app:GKW}. Note in particular that $m_{3/2}$ is used as the ``apparent gravitino mass'', which appears in the Lagrangian.

\subsection{Lagrangian with Goldstino}

In the Lagrangian (\ref{SugraLagr}) there is a mixing term between the gravitino and the Goldstino $\upsilon$. In order to `diagonalize' the Lagrangian, we redefine the gravitino with terms proportional to the Goldstino. The procedure has been explained in detail in \cite[Sec.19.1.1]{Freedman:2012zz} for Minkowski space. However, it will turn out that in curved space more terms are generated even for time-independent configurations of the fields. Therefore we consider a redefinition of the gravitino that corresponds to a supersymmetry transformation of the gravitino with parameter $\epsilon = \alpha \upsilon $, where $\alpha $ will be determined by the requirement of absence of mixing terms between the gravitino and the other fermions:
\begin{equation}
  P_L\psi _\mu =P_L\Psi_\mu +\alpha D _\mu P_L\upsilon +\ft12\alpha m_{3/2}\gamma _\mu P_R\upsilon \,.
\label{massivegravitino}
\end{equation}
Considering first the gravitino kinetic terms (here and below, we omit the overall factor $e$):
\begin{align}
& -\ft1{2}\bar \psi_\mu \gamma ^{\mu \rho \sigma }D _{\rho }\psi _\sigma\nonumber\\
    & = -\ft1{2}\left[\overline{\Psi}_\mu +\alpha D _\mu \overline{\upsilon }-\ft12\alpha \overline {m}_{3/2} \bar \upsilon \gamma _\mu\right]P_R\gamma ^{\mu \rho \sigma }D _{\rho }
    \left[\Psi_\sigma  +\alpha D_\sigma  \upsilon +\ft12\alpha m_{3/2}\gamma _\sigma \upsilon\right]+\hc\nonumber\\
    &=-\ft1{2}\bar \Psi_\mu \gamma ^{\mu \rho \sigma }D _{\rho }\Psi _\sigma -\alpha \overline{\Psi}_\mu \gamma ^{\mu \rho \sigma }D _{\rho }D_\sigma  \upsilon
    -\alpha \overline{\Psi}_\mu(P_Rm_{3/2}+P_L\overline{m}_{3/2})\gamma ^{\mu \nu }D_\nu \upsilon \nonumber\\
    & + \alpha ^2 \overline{\upsilon }\left[\ft12\gamma ^{\mu \rho \sigma }D_\mu D_\rho D_\sigma
     +(P_Rm_{3/2}+P_L\overline{m}_{3/2})  \gamma ^{\mu \nu  }D_\mu D _{\nu }
     +\ft34|m_{3/2}|^2 \slashed{D} \right]\upsilon  \,,
\label{gravitinokinredef}
\end{align}
where we omitted total derivatives, neglected higher order terms due to torsion, \ldots .

Similar for the mass term
\begin{align}
 & \ft12 m_{3/2}\bar \psi _\mu P_R
\gamma ^{\mu \nu }\psi _\nu +\hc \nonumber\\
&= \ft12 m_{3/2}\left[\overline{\Psi}_\mu +\alpha D _\mu \overline{\upsilon }-\ft12\alpha \bar {m}_{3/2} \bar \upsilon \gamma _\mu\right]P_R\gamma ^{\mu \nu }
\left[\Psi_\nu +\alpha D _\nu \upsilon +\ft12\alpha \overline{m}_{3/2}\gamma _\nu \upsilon\right]+\hc\nonumber\\
&= \ft12\bar \Psi _\mu (P_Rm_{3/2}+P_L\overline{m}_{3/2}) \gamma ^{\mu \nu }\Psi _\nu
+ \alpha  \overline{\Psi}_\mu (P_Rm_{3/2}+P_L\overline{m}_{3/2})  \gamma ^{\mu \nu }D _\nu \upsilon +\ft32\alpha |m_{3/2}|^2\overline{\Psi}\cdot \gamma\upsilon\nonumber\\
&+\alpha ^2\overline{\upsilon }\left[ -\ft12 (P_Rm_{3/2}+P_L\overline{m}_{3/2}) \gamma ^{\mu \nu }D_\mu D_\nu
-\ft32|m_{3/2}|^2\slashed{D} -\ft32|m_{3/2}|^2(P_Rm_{3/2}+P_L\overline{m}_{3/2})\right]\upsilon \,.
 \label{mgravitredef}
\end{align}

Finally, from the mixing term
\begin{align}
  &-  \overline{\psi}  \cdot \gamma P_L\upsilon +\hc\nonumber\\
  &= - \left[\overline{\Psi}_\mu +\alpha D _\mu \overline{\upsilon }-\ft12\alpha \overline{m}_{3/2} \bar \upsilon \gamma _\mu\right]  \cdot \gamma^\mu  P_L\upsilon+ \hc\nonumber\\
  &= -  \overline{\Psi}  \cdot \gamma \upsilon + \alpha \overline{\upsilon } \slashed{D}\upsilon +2\alpha \overline{\upsilon }(P_Rm_{3/2}+P_L\overline{m}_{3/2})\upsilon \,.
 \label{mixredef}
\end{align}

The terms linear in $\alpha $ are those produced by a supersymmetry transformation. If also the frame field is redefined as in a supersymmetry transformation,
\begin{equation}
  e_\mu {}^a = E_\mu {}^a  +\ft12\alpha \bar \upsilon \gamma ^a\Psi _\mu \,,
 \label{redefframe}
\end{equation}
the terms proportional to $\alpha $ should cancel for the supersymmetric anti-de Sitter case where $V= 3|m_{3/2}|^2$, and there is no `mixing term'.
Indeed, the redefinition (\ref{redefframe}) leads to terms proportional to the Einstein tensor and terms due to the potential term:
\begin{align}
  \ft12 e R(e)-e\, V &=  \ft12 E R(E)-E\, V - E\left[R_a{}^\mu(E) -\ft12 R(E)E_a{}^\mu+V\, E_a{}^\mu\right]\ft12\alpha \bar \upsilon \gamma ^a\Psi _\mu +\ldots\nonumber\\
&= \ft12 E R(E)-E\, V + \ft12\alpha E\left[\left( R_{\mu a}(E)-\ft12E_{\mu a}R(E)\right)\overline{\Psi}^a \gamma ^\mu \upsilon +  V\,  \bar \Psi\cdot \gamma \upsilon\right]+\ldots \,.
 \label{frameredef}
\end{align}
There are many terms in higher orders in the fermions, which we do not have to consider.

We then obtain as terms linear in the Goldstino
\begin{equation}
  \left[\ft32\alpha |m_{3/2}|^2 +\ft12 \alpha V-1\right]    \overline{\Psi}\cdot \gamma\upsilon
 \label{linupsilon}
\end{equation}
Indeed, the $\alpha $ terms cancel for the supersymmetric anti-de Sitter case. We fix $\alpha $ by the requirement that such mixing terms do not occur, and thus obtain:
\begin{equation}
  \alpha = \frac{2}{V+3|m_{3/2}|^2}=\frac{2}{V_+}= \frac{2}{|m_{3/2}|^2\, X}\,,
 \label{valuealpha}
\end{equation}
where $V_+$ and $X$ are defined in (\ref{potN1sugrafinal})--(\ref{defXG}).

There remain four types of terms quadratic in the Goldstino:
\begin{enumerate}
  \item Kinetic terms
\begin{equation}
  \left(-\ft34\alpha ^2|m_{3/2}|^2+\alpha \right)\overline{\upsilon }\slashed{D}\upsilon \,.
 \label{kinetic}
\end{equation}
  \item Mass-like terms
\begin{equation}
m_{3/2}\left(- \ft32\alpha ^2|m_{3/2}|^2+2\alpha \right)\overline{\upsilon }P_R\upsilon +\hc\,.
 \label{masslike}
\end{equation}
\item A term with a triple covariant derivative
\begin{equation}
  \ft12\alpha ^2 \overline{\upsilon }\gamma ^{\mu \rho \sigma }D_\mu D_\rho D_\sigma\upsilon \,.
 \label{tripleD}
\end{equation}
  \item  Terms which lead to a commutator of Lorentz-covariant derivatives
\begin{equation}
  \ft12m_{3/2} \alpha ^2\overline{\upsilon }P_R \gamma ^{\mu \nu }D_\mu D_\nu \upsilon +\hc\,.
 \label{commutatorterms}
\end{equation}
\end{enumerate}
For the latter two terms, we can use the same methods as in the proof of the invariance of the supergravity action\footnote{Details can be found in\cite[Sec.9.1]{Freedman:2012zz}.}
\begin{align}
  [D_\mu  ,\,D_\nu  ]\upsilon =\ft14 R_{\mu \nu  }{}^{ab}\gamma _{ab}\upsilon \qquad \rightarrow \qquad &\gamma ^{\mu \rho \sigma}D_\rho D_\sigma\upsilon = \ft12 R^{\mu \nu }\gamma _\nu \upsilon -\ft14\gamma ^\mu R\upsilon\,,\nonumber\\
  &\gamma ^{\mu \nu }D_\mu D_\nu \upsilon=-\ft14 R\upsilon \,.
 \label{commutatorD}
\end{align}
Using also the Bianchi identity, these terms therefore lead to
\begin{align}
  3:\qquad  & \ft14\alpha ^2 \overline{\upsilon }\left(R^{\mu \nu }\gamma _\nu -\ft12\gamma ^\mu R\right)D_\mu \upsilon \,,\nonumber\\
  4:\qquad & -\ft18\alpha ^2R\, \overline{\upsilon }(P_Rm_{3/2}+P_L\overline{m}_{3/2})\upsilon\,.
\label{curvatureterms}
\end{align}

To cancel these terms, we can modify the frame field once more:
\begin{equation}
  E_\mu {}^a = E'_\mu {}^a +  \ft18\alpha ^2\overline{\upsilon }\left[2\gamma ^a D_\mu +E'_\mu {}^a (P_Rm_{3/2}+P_L\overline{m}_{3/2})\right]\upsilon\,.
 \label{EEprime}
\end{equation}
Similar to (\ref{frameredef}), this leads to extra terms in the action (rewriting now the new $E'_\mu {}^a$ as conventional frame field $e_\mu {}^a$)
\begin{align}
&-\ft18\alpha ^2e\,\left(R_a{}^\mu -\ft12 R e_a{}^\mu +Ve_a{}^\mu\right)\overline{\upsilon }\left[2\gamma ^a D_\mu +e_\mu {}^a (P_Rm_{3/2}+P_L\overline{m}_{3/2})\right]\upsilon\nonumber\\
=&-\ft14\alpha ^2e\,\overline{\upsilon }\left(R^{\mu \nu }\gamma _\nu -\ft12\gamma ^\mu R\right)D_\mu \upsilon +
\ft18 e\, R\,\overline{\upsilon }(P_Rm_{3/2}+P_L\overline{m}_{3/2})\upsilon\,.\nonumber\\
&-\ft14\alpha ^2e\,V\,\overline{\upsilon }\slashed{D}\upsilon
-\ft12e\,\alpha ^2\,V\,\overline{\upsilon }(P_Rm_{3/2}+P_L\overline{m}_{3/2})\upsilon\,.
 \label{extrafromEprime}
\end{align}
The first line of the right-hand side cancels the terms in (\ref{curvatureterms}), while the last line gives new terms of the form of  (\ref{kinetic}) and (\ref{masslike}).

Therefore, we remain after the super-BEH mechanism with the new terms that are proportional to
\begin{equation}
  -\ft32\alpha ^2|m_{3/2}|^2+2\alpha -\ft12 \alpha ^2 V = \alpha \,,
 \label{propalpha}
\end{equation}
using (\ref{valuealpha}).

The upshot is that this super-BEH mechanism leads to  terms
\begin{equation}
e^{-1}{\cal L}^{(\upsilon )}= \ft12\alpha \,\overline{\upsilon }\slashed{D}\upsilon+  \alpha\,\overline{\upsilon }(P_Rm_{3/2}+P_L\overline{m}_{3/2})\upsilon\,.
 \label{Lupsilon}
\end{equation}
Using the value of the Goldstino as in (\ref{defnquant}), these terms are of the form (using the quantity $X$ defined in (\ref{defX}) and neglecting spacetime derivatives on the scalars)
\begin{align}
  e^{-1}{\cal L}^{(\upsilon )}&=\frac{\rme^{{\cal K}}}{|m_{3/2}|^2\,X}\left(\chi ^\alpha \slashed{D}\chi ^{\bar \beta }\nabla _\alpha W \overline{\nabla }_{\bar \beta }\overline{W} +\overline{m}_{3/2}\bar \chi ^\alpha \chi ^\beta  \nabla _\alpha W\nabla _\beta W+{m}_{3/2}\bar \chi ^{\bar \alpha }\chi ^{\bar \beta }\overline{\nabla }_{\bar \alpha }W\overline{\nabla }_{\bar \beta }\overline{W}\right)\nonumber\\
  &=\frac{1}{X}\left(\chi ^\alpha \slashed{D}\chi ^{\bar \beta }{\cal G} _\alpha\overline{{\cal G}}_{\bar \beta } +\rme^{{\cal G}/2}\sqrt{\frac{W}{\overline{W}}}\bar \chi ^\alpha \chi ^\beta  {\cal G}_\alpha {\cal G} _\beta +\rme^{{\cal G}/2}\sqrt{\frac{\overline{W}}{W}}\bar \chi ^{\bar \alpha }\chi ^{\bar \beta }\overline{{\cal G} }_{\bar \alpha }\overline{{\cal G} }_{\bar \beta }\right)\,.
 \label{Lupschi}
\end{align}
The K\"{a}hler-invariant function ${\cal G}$ is defined in (\ref{calGdef}).

\subsection{Modified kinetic terms}

Combining the kinetic terms of the fermions of (\ref{SugraLagr}) with the correction terms in (\ref{Lupschi}) leads to a kinetic matrix
\begin{align}
  G_{\alpha \bar \beta }&= g_{\alpha \bar \beta }-\frac{\rme^{{\cal K}}}{|m_{3/2}|^2\,X}\nabla _\alpha W \overline{\nabla }_{\bar \beta }\overline{W}\nonumber\\
  &=g_{\alpha \bar \beta }-\frac{1}{X}{\cal G} _\alpha \overline{{\cal G} }_{\bar \beta }\,.
 \label{modifKin}
\end{align}
This has the zero mode
\begin{equation}
  G_{\alpha \bar \beta }\nabla ^{\bar \beta }W = G_{\alpha \bar \beta }{\cal G}^{\bar \beta }=0\,.
 \label{zeromodekin}
\end{equation}
This proves that the Goldstino disappears from the kinetic matrix.

\subsection{Modified mass terms}

In the same way, the full holomorphic mass matrix is
\begin{equation}
 M_{\alpha \beta }= m_{\alpha \beta } - \frac{2}{m_{3/2}X}\rme^{{\cal K}}\nabla _\alpha W\nabla _\beta W= \sqrt{\frac{W}{\overline{W}}}
 \rme^{{\cal G}/2}\left[{\cal G}_{\alpha \beta }+ \frac{X-2}{X}{\cal G}_\alpha {\cal G}_\beta\right]\,.
 \label{fullM}
\end{equation}
Note that, with the help of (\ref{dalphaV}), we can derive that this mass term has a zero mode proportional to the Goldstino modulo the derivative of the potential:
\begin{equation}
 M_{\alpha \beta }\overline{\nabla }^\beta \overline{W}= \rme^{-{\cal K}/2}  V_\alpha\,,\qquad
  M_{\alpha \beta }\overline{{\cal G}}^\beta = \sqrt{\frac{W}{\overline{W}}}\rme^{-{\cal G}/2} V_\alpha\,,
 \label{zeromode}
\end{equation}
so that for $V_\alpha =0$ the matrix $M$ has rank $N-1$.

One could also define an off-shell mass matrix (with $V_\alpha =\partial _\alpha V$)
\begin{equation}
  M_{\alpha \beta }^{\rm OS}= M_{\alpha \beta }- \sqrt{\frac{W}{\overline{W}}}\rme^{-{\cal G}}\frac{V_\alpha V_\beta }{V_\gamma \overline{{\cal G}}^\gamma }\,,
 \label{MOS}
\end{equation}
which satisfies
\begin{equation}
  M_{\alpha \beta }^{\rm OS}\overline{{\cal G}}^\beta = 0\,.
 \label{zeromodeOS}
\end{equation}

\section{Supertrace formulae}
\label{ss:Str}

\subsection{Definitions of supertrace for broken and unbroken supersymmetry}

For unbroken supersymmetry the supertrace depends only on the scalar and spin 1/2 masses. It takes the form
\begin{eqnarray}
\ft12 \mathop{\rm Supertr}\nolimits{\cal M}^2&\equiv&\ft12\sum_{J}(-)^{2J}(2J+1)m_J^2\nonumber\\
 &=&\ft12\tr {\cal M}_{0}^2-\tr {\cal M}_{1/2}^2\nonumber\\
 &=& g^{\alpha \bar \beta }V_{\alpha \bar \beta }-m_{\alpha \beta }\overline{m}^{\alpha \beta }\,.
 \label{defsTrunbroken}
\end{eqnarray}
For broken supersymmetry, also the gravitino enters, and the Goldstino modifies the mass matrix of the spin 1/2 particles. Hence
\begin{align}
\ft12 \mathop{\rm Supertr}\nolimits{\cal M}^2 &=\ft12\tr {\cal M}_{0}^2-\tr {\cal M}_{1/2}^2- 2|m_{3/2}|^2\nonumber\\
 &= g^{\alpha \bar \beta }V_{\alpha \bar \beta }-M_{\alpha \beta }\overline{M}^{\alpha \beta }- 2|m_{3/2}|^2\,.
 \label{defsTrbroken}
\end{align}
Note that by masses we mean the Lagrangian masses. In Sec. \ref{ss:conclusion} we will comment on other definitions of the mass.

\subsection{Rigid supersymmetry}
For pedagogical purpose, we first make the calculation for rigid supersymmetry in Minkowski space. The masses for the scalars are determined by the second derivatives of the potential\footnote{$W_\alpha =\partial _\alpha W$, indices are raised with the metric such that $\overline{W}^\alpha =g^{\alpha \bar \beta }\overline{W}_\beta $ and covariant derivatives use only the Christoffel symbols of this metric.}
\begin{align}
  V =& W_\alpha \,\overline{W}^\alpha = W_\alpha g^{\alpha \bar \beta }\overline{W}_{\bar \beta } \,, \nonumber\\
  V_\alpha =&\partial _\alpha V = \nabla _\alpha  W_\beta  \,\overline{W}^\beta \,,\nonumber\\
  V_{\alpha \bar \alpha }=&\partial _\alpha \partial _{\bar \alpha } V = m_{\alpha \beta  }g^{\beta  \bar \beta  }\overline{m}_{\bar \alpha \bar \beta }+ R_{ \alpha \bar\alpha }{}^{ \beta\bar\beta } W_\beta \,\overline{W}_{\bar \beta}\,,\nonumber\\
  V_{\alpha \beta }=&\nabla _\alpha \partial _\beta V=\nabla _\alpha\nabla _\beta W_\gamma \, \overline{W}^\gamma \,,
\label{Vrigid}
\end{align}
where for rigid supersymmetry
\begin{equation}
  m_{\alpha \beta }= \nabla _\alpha \partial _\beta W\,.
 \label{mrigid}
\end{equation}
The latter is the holomorphic fermion mass matrix.
We used commutators of covariant derivatives like
\begin{equation}
  \left[\nabla_\alpha,\overline{\nabla}_{\bar{\beta}}\right]W_\gamma =R_{\alpha\bar{\beta}\gamma}{}^\delta W_\delta\,,\qquad
  R_{\alpha\bar{\beta}}\,=\,g^{\bar\gamma \gamma  }R_{\alpha \bar \gamma \bar \beta \gamma  }=
-R_{\alpha \bar \beta  \gamma  }{}^\gamma\,.
 \label{defcurv}
\end{equation}
We mention in (\ref{Vrigid}) also the holomorphic second derivative, despite the fact that it is not relevant for the trace of the mass matrix. However, it governs a mass splitting between the real and imaginary parts of the complex bosons if supersymmetry is broken.

Supersymmetry is unbroken if
\begin{equation}
  \mbox{Unbroken supersymmetry: }\qquad W_\alpha =\overline{W}_{\bar \alpha }=0\,.
 \label{unbrokenSUSY}
\end{equation}
In that case, we see that we have always an extremum of the potential $V_\alpha=0$ where the potential is zero, $V=0$. Since then also the last line of (\ref{Vrigid}) vanishes, and $V_{\alpha \bar \alpha }$ reduces to the square of the fermion mass matrix, these scalars and fermions for each $\alpha $ have the same mass in a basis where the metric is diagonalized.

The supertrace of the mass matrix is
\begin{equation}
  \ft12 \mathop{\rm Supertr}\nolimits{\cal M}^2 =\ft12\tr {\cal M}_{0}^2-\tr {\cal M}_{1/2}^2= g^{\alpha \bar \beta }V_{\alpha \bar \beta }-m_{\alpha \beta }\overline{m}^{\alpha \beta }=
   R^{\bar \beta\beta } W_\beta \,\overline{W}_{\bar \beta}\,.
 \label{supertraceRigid}
\end{equation}
In \cite[Sec.14.5.5]{Freedman:2012zz} this is explained in more detail, and the generalisation including gauge multiplets is given.

\subsection{Calculation of the supergravity trace formula}

For the bosonic masses we need the second derivatives of the potential. These are modifications of (\ref{Vrigid}), starting with the potential in (\ref{potN1sugrafinal}) or  (\ref{VGX}) and first derivatives in (\ref{dalphaV}). Note that the conditions of unbroken supersymmetry
\begin{equation}
  \mbox{Unbroken supersymmetry: }\qquad \nabla _\alpha  W =0\,,\qquad \mbox{or}\qquad {\cal G}_\alpha =0\,,
 \label{unbrokenSUGRA}
\end{equation}
imply that to the supersymmetry-preserving conditions lead to an extremum of the potential. The first derivative of the potential, $V_\alpha $, is given by (\ref{dalphaV}).

The second derivatives (in both formulations) are
\begin{align}
V_{\alpha\bar \alpha }=&(V+|m_{3/2}|^2)g_{\alpha \bar \alpha } +m_{\alpha \beta }g^{\beta \bar \beta }\overline{m}_{\bar \alpha \bar \beta }+\rme^{{\cal K}}\left(-
\nabla _\alpha W\overline{\nabla }_{\bar \alpha }\overline{W}+R_{\alpha \bar \alpha} {}^{\beta\bar  \beta } \nabla _\beta  W\,\overline{\nabla }_{\bar \beta}\overline{W}\right)
\nonumber\\
=&\rme^{{\cal G}}\left[g_{\alpha \bar \alpha }(X-2)-{\cal G}_\alpha \overline{{\cal G}_{\bar \alpha }}+\left({\cal G}_{\alpha \beta }+{\cal G}_\alpha {\cal G}_\beta \right)g^{\beta \bar \beta }\left( \overline{{\cal G}}_{\bar \alpha \bar \beta }+\overline{{\cal G}}_{\bar \alpha }\overline{{\cal G}}_{\bar \beta }\right) + R_{\alpha \bar \alpha} {}^{\beta\bar  \beta } {\cal G} _\beta  \,\overline{{\cal G} }_{\bar \beta}\right]\,,\nonumber\\
V_{\alpha \beta }=&\rme^{{\cal K}}\left[ -\overline{W}\nabla _\alpha \nabla _\beta W +\overline{\nabla }^\gamma \overline{W}\,\nabla _\alpha \nabla _\beta \nabla _\gamma W\right] \nonumber\\
=&\rme^{{\cal G}}\left[-\left({\cal G}_{\alpha \beta }+{\cal G}_\alpha {\cal G}_\beta \right) +\left( {\cal G}_{\alpha \beta \gamma }+{\cal G}_\alpha {\cal G}_\beta {\cal G}_\gamma +3{\cal G}_{(\alpha \beta }{\cal G}_{\gamma )}\right) \overline{{\cal G} }^\gamma\right] \,.
\label{Vderiv}
\end{align}
In the unbroken case, they reduce to
\begin{eqnarray}
 V_{\alpha\bar \alpha }  & =&-2 |m_{3/2}|^2g_{\alpha \bar \alpha }+ m_{\alpha \beta }g^{\beta \bar \beta }\overline{m}_{\bar \alpha \bar \beta }= \rme^{{\cal G}}\left[-2 g_{\alpha \bar \alpha }+{\cal G}_{\alpha \beta }g^{\beta \bar \beta }\overline{{\cal G}}_{\bar \alpha \bar \beta }\right] \,, \nonumber\\
 V_{\alpha \beta } &=&  -\overline{m}_{3/2}m_{\alpha \beta }= -\rme^{{\cal G}}{\cal G}_{\alpha \beta }\,.
 \label{Vderivunbroken}
\end{eqnarray}
For the trace of the mass matrix, only the mixed derivatives contribute, and we have
\begin{align}
  \ft12\tr {\cal M}_{0}^2
  =&-2N|m_{3/2}|^2+ m_{\alpha \beta }\overline{m}^{\alpha \beta }+  \rme^{{\cal K}} \left[(N-1)
  \nabla _\alpha W\overline{\nabla}^\alpha  \overline{W}
  +R^{\bar \beta \beta  }\nabla _\beta  W\overline{\nabla }_{\bar \beta }\overline{W}
\right]\nonumber\\
=&\rme^{{\cal G}}\left[ (N-1){\cal G}_\alpha \overline{{\cal G}}^\alpha -2N+ \left( {\cal G}_{\alpha \beta }+{\cal G}_\alpha {\cal G}_\beta \right) \left(\overline{{\cal G}}^{\alpha \beta }+\overline{{\cal G}}^\alpha \overline{{\cal G}}^\beta \right) +{\cal G}_\alpha\overline{{\cal G}}_{\bar \beta}R^{\bar \beta \alpha }\right] \,.
 \label{trM02}
\end{align}

Using the zero-mode equation (\ref{zeromode}), a useful relation can be derived for the square of the fermion mass matrix:
\begin{align}
  m_{\alpha \beta }\overline{m}^{\alpha \beta}=& M_{\alpha \beta }\overline{M}^{\alpha \beta}+ 4|m_{3/2}|^2 +\frac{2}{X}\left( \frac{1}{\overline{W}}V_\alpha \overline{\nabla }^\alpha \overline{W}+ \frac{1}{W}V_{\bar\alpha}\nabla ^{\bar \alpha }W\right) \nonumber\\
  =& M_{\alpha \beta }\overline{M}^{\alpha \beta}+ 4\rme^{{\cal G}} +\frac{2}{X}\left( V_\alpha \overline{{\cal G}}^\alpha +V_{\bar\alpha} {\cal G}^{\bar\alpha} \right)\,.
 \label{m2toM2}
\end{align}

This reduces the previous formula to
\begin{align}
  \ft12\tr {\cal M}_{0}^2=&(N-1) V + (N+1)|m_{3/2}|^2+ M_{\alpha \beta }\overline{M}^{\alpha \beta }  +  \rme^{{\cal K}} R^{\bar \beta \beta  }\nabla _\beta  W\overline{\nabla }_{\bar \beta }\overline{W}\nonumber\\ &+\frac{2}{X}\left( V_\alpha \overline{{\cal G}}^\alpha+V_{\bar\alpha} {\cal G}^{\bar\alpha}\right)
 \label{trM0simple}
\end{align}

\subsection{Results}

The supertrace defined by (\ref{defsTrbroken}) is therefore for $N$ chiral multiplets
\begin{align}
   \ft12 \mathop{\rm Supertr}\nolimits{\cal M}^2 =& (N-1) (V+|m_{3/2}|^2) + \rme^{{\cal K}}\overline{\nabla }^\alpha \overline{W}\,\nabla ^{\bar \beta }W\,R_{\alpha  \bar \beta }\nonumber\\
   &+   \frac{\rme^{\cal K}}{V_+}\left(V_\alpha \, W\,\overline{\nabla }^\alpha\overline{W} +V_{\bar \alpha}\, \overline{W}\,\nabla ^{\bar \alpha}W\right)\,,\nonumber\\
   \ft12 \mathop{\rm Supertr}\nolimits{\cal M}^2 =& (N-1) (V+|m_{3/2}|^2) + \rme^{{\cal G}}\overline{{\cal G}}^\alpha {\cal G}^{\bar \beta }\,R_{\alpha  \bar \beta }\nonumber\\
  &+   \frac{\rme^{\cal G}}{V_+}\left(V_\alpha \overline{{\cal G}}^\alpha +V_{\bar \alpha} {\cal G}^{\bar \alpha}\right)\,.
 \label{resultSTrG}
\end{align}
As mentioned in (\ref{potN1sugrafinal}), $V_+= V+3|m_{3/2}|^2$.
Note that the last term is singular for unbroken supersymmetry, since then $V_+=0$. At an extremum $V_\alpha =0$ we obtain
\begin{equation}
  \ft12 \mathop{\rm Supertr}\nolimits{\cal M}^2 = (N-1) (V+|m_{3/2}|^2) + \rme^{{\cal K}}\overline{\nabla }^\alpha \overline{W}\,\nabla ^{\bar \beta }W\,R_{\alpha  \bar \beta }\,,
 \label{STrExtremum}
\end{equation}
which is the generalization of the result in \cite{Cremmer:1982en,Grisaru:1982sr} for arbitrary cosmological constant $V$.  Note that for $N=1$ the first term drops for arbitrary $V$.

A particularly interesting situation is the configuration for which $V_-=0$ and $V_+$ is non-zero, which implies a
positive cosmological constant with vanishing `apparent gravitino mass' $m_{3/2}$. This is a realisation of the
Super-BEH effect in de Sitter space. Also in  this case, looking at  (\ref{resultSTrG}), there is a breaking term $(N-1)V_+$, and a K\"{a}hler curvature term while the last term vanishes.

In the case of unbroken supersymmetry, see (\ref{unbrokenSUGRA}), in AdS space, we use (\ref{defsTrunbroken}) to obtain the mass formula for chiral multiplets in AdS. The result (\ref{Vderivunbroken}) then gives
\begin{equation}
  \ft12 \mathop{\rm Supertr}\nolimits{\cal M}^2 =-2N\,|m_{3/2}|^2= -2N\,\rme^{{\cal G}}\,.
 \label{unbrokensTr}
\end{equation}

The masses for $N=1$ (one chiral multiplet) for unbroken supersymmetry in AdS were obtained in \cite{Ferrara:2013pla}. In  \cite{Ferrara:2016vzg} the formula for $V$ and its derivative $V'$ not zero is given for a single chiral multiplet, which therefore must be the sgoldstino.

For unbroken supersymmetry in Minkowski, the supertrace vanishes completely, and one has mass degeneracy of all the spin 0 and spin 1/2 particles.

\section{Concluding remarks}
\label{ss:conclusion}

In this work we gave the generalisations of the trace formulae of  \cite{Ferrara:1979wa,Cremmer:1982en,Grisaru:1982sr} to curved spacetime
where configurations with non-vanishing potential and also with non-vanishing first derivative are considered.
These formulae include the case of unbroken supersymmetry in Minkowski and AdS backgrounds
as well as the case with broken supersymmetry with arbitrary sign of the cosmological constant.

We emphasize that our formulae are based on the mass parameters in the Lagrangian. One may consider different definitions of the mass in curved space. E.g. for  spin-1/2 fermions the equation of motion derived from the action is
\begin{equation}
  \slashed{\nabla}  \chi + m\chi =0\,,
 \label{chimassnaive}
\end{equation}
where $m$ is the  parameter in the Lagrangian term $-\ft12m\bar \chi \chi$.
Squaring this equation in curved space using (\ref{commutatorD}) give
\begin{align}
 0= \slashed{\nabla }\slashed{\nabla }\chi  -m^2\chi  =\Box \chi  -\ft14 R \chi -m^2\chi \,.
 \label{quadrm12}
\end{align}
This might be interpreted as a squared mass of the fermion equal to $m^2+\ft14 R$. However, since we do not know of any general accepted definition of masses in (anti-) de Sitter space, we formulated our results in terms of the mass parameters in the Lagrangian.

Configurations with $W=0$ imply either unbroken supersymmetry in flat space or a de Sitter geometry with
broken supersymmetry. Otherwise $W$ non-zero is the general situation that also includes broken supersymmetry in flat space and unbroken supersymmetry in AdS space.
The formulae for rigid (global) supersymmetry were also given and shown to be related to the local formulae in particular limits.
We also retrieved the difference between mass relations in flat and AdS supersymmetric backgrounds.

An alternative supertrace formula can be written down using the concept of an `effective squared gravitino mass'
\begin{equation}
  m_{3/2}^{2\ \rm eff}=|m_{3/2}|^2+\ft13V\,.
 \label{m32eff}
\end{equation}
This expression has the property of vanishing whenever supersymmetry is unbroken.
An effective supertrace is then defined as
\begin{equation}
  \ft12 \mathop{\rm Supertr}\nolimits{\cal M}^2_{\rm eff} \equiv\ft12\tr {\cal M}_{0}^2-\tr {\cal M}_{1/2}^2-2 m_{3/2}^{2\ \rm eff}
  =\ft12 \mathop{\rm Supertr}\nolimits{\cal M}^2-\ft23 V\,.
 \label{defStreff}
\end{equation}
Using the effective gravitino mass also in the right-hand side of (\ref{STrExtremum}), we obtain
\begin{equation}
  \ft12 \mathop{\rm Supertr}\nolimits{\cal M}^2_{\rm eff}+\ft43 V = (N-1) m_{3/2}^{2\ \rm eff}+\ft23 N\,V + \rme^{{\cal K}}\overline{\nabla }^\alpha \overline{W}\,\nabla ^{\bar \beta }W\,R_{\alpha  \bar \beta }
 \label{effstrformula}
\end{equation}
For unbroken supersymmetry, the effective gravitino mass vanishes and with $V= -3|m_{3/2}|^2$,
the left- and right-hand side go smoothly into (\ref{unbrokensTr}).

These results are valid for spontaneously broken linearly realised ${\cal N}=1$ supersymmetry with multiplets where the Goldstino has always a scalar partner, the sgoldstino. However, one may envisage linear theories where in some limits the sgoldstino has a large mass with respect to supersymmetry breaking scale \cite{Komargodski:2009rz,Kallosh:2016hcm}  and can be integrated out giving a nonlinear realisation of supersymmetry of the Volkov-Akulov type \cite{Volkov:1973ix,Casalbuoni:1988xh,Komargodski:2009rz,Bergshoeff:2015tra}.

In this limiting case it should be possible to adapt the supertrace formulae to nonlinear realisations
as is the case for the scalar potential, which has been largely explored in recent time especially in relation to models of inflation  \cite{Antoniadis:2014oya,Ferrara:2014kva,Kallosh:2014via,Dall'Agata:2014oka}\footnote{We only included early references on the subject. For a recent review, see \cite{Ferrara:2016ajl}.} and the KKLT \cite{Kachru:2003aw} scenario.

%%%%%%%%%%%%%%%%%%%%%%%%%%%%%%%%
\medskip
\section*{Acknowledgments.}

\noindent We thank E. Bergshoeff, G. Dall'Agata, D. Freedman, R. Kallosh, A. Linde, D. Roest and A. Sagnotti for discussions and collaborations on related topics.

We acknowledge hospitality of the GGI institute in Firenze, where this work was performed during the workshop `Supergravity: what next?'.
This work is supported in part by the COST Action MP1210 `The
String Theory Universe'.

The work of SF is supported in part by CERN TH Dept and INFN-CSN4-GSS.
The work of A.V.P. is supported in part by the Interuniversity Attraction Poles Programme
initiated by the Belgian Science Policy (P7/37).

\bigskip

%\newpage
%%%%%%%%%%%%%%%%%%%%%%%%%%%
\appendix
\section{Notation and translation between two formalisms}
%from formulaes in terms of the K\"{a}hler-invariant function \texorpdfstring{${\cal G}$}{G}.}
\label{app:GKW}

We put the gravitational coupling constant $\kappa =1$.  The K\"{a}hler-invariant function is determined by the K\"{a}hler potential ${\cal K}$ and the superpotential $W$ by
\begin{equation}
{\cal G}={\cal K} +\log (W\overline W) \,,
 \label{calGdef}
\end{equation}
Simple derivatives lead to relations with the K\"{a}hler covariant derivatives of the superpotential:
\begin{align}
  {\cal G}_\alpha &\equiv \partial _\alpha {\cal G}  = \frac{1}{W}\nabla _\alpha W\,, \nonumber\\
  {\cal G}_{\alpha \beta } & \equiv \partial _\alpha {\cal G}_\beta -\Gamma _{\alpha \beta }^\gamma {\cal G}_\gamma  = \frac{1}{W}\nabla _\alpha\nabla _\beta  W -{\cal G}_\alpha {\cal G}_\beta \,,
\label{Gderiv}
\end{align}
where
\begin{equation}
  \nabla _\alpha W(z) =\partial _\alpha W(z) +  (\partial_\alpha {\cal K})   W(z)\,.
\label{nablaWfinal}
\end{equation}
The third derivatives are related by
\begin{equation}
  \rme^{{\cal K}/2}\nabla _\alpha \nabla _\beta \nabla _\gamma W = \sqrt{\frac{W}{\overline{W}}}\rme^{{\cal G}/2}\left({\cal G}_{\alpha \beta \gamma }+{\cal G}_\alpha {\cal G}_\beta {\cal G}_\gamma+ 3{\cal G}_{(\alpha \beta }{\cal G}_{\gamma )}\right)\,.
 \label{3rdderivrelation}
\end{equation}

Main quadratic terms of the supergravity Lagrangian are
\begin{align}
  e^{-1}{\cal L} =& \frac1{2}\left[ R(e)-\bar \psi_\mu \gamma ^{\mu \rho \sigma }D _{\rho }\psi _\sigma \right]
  -g_{\alpha \bar \beta }\left[\partial _\mu z^\alpha
\partial^\mu\bar z^{\bar \beta }
+\,\bar \chi ^\alpha {\slashed{D}}\chi^{\bar
\beta }\right]-V  \nonumber\\
    & +\left[\ft12 m_{3/2}\bar \psi _\mu P_R
\gamma ^{\mu \nu }\psi _\nu  -\,\ft12 m_{\alpha \beta }\bar
\chi^\alpha \chi^\beta
%-m_{\alpha A}\bar \chichiral^\alpha \PLa\lambda ^A-\ft12m_{AB}\bar \lambda ^AP_L\lambda^B
-\bar  \psi  \cdot \gamma P_L\upsilon +\hc\right]+\ldots
\label{SugraLagr}
\end{align}
where $\chi ^\alpha $ are the left-handed fermions ($P_L\chi ^\alpha =\chi ^\alpha $) and $\chi ^{\bar \alpha }$ are right-handed. For this paper it is sufficient to consider $D_\mu $ as the Lorentz-covariant derivative (other terms are higher order in the fields). Other notations in (\ref{SugraLagr}) are
\begin{align}
   m_{3/2}=& \rme^{{\cal K}/2}W\,,
 \label{m32N1}\\
  m_{\alpha \beta }  =& \rme^{{\cal K}/2}\nabla _\alpha \nabla _\beta W\,, \nonumber\\
  P_L\upsilon  & =-\frac{1}{\sqrt{2}}\chi ^\alpha  \rme^{{\cal K}/2} \nabla _\alpha W\,.
\label{defnquant}
\end{align}
In the ${\cal G}$-formalism,
\begin{align}
  |m_{3/2}|^2 =& \rme^{\cal G} \nonumber\\
  m_{\alpha \beta }  = & \sqrt{\frac{W}{\overline{W}}}\rme^{{\cal G}/2}\left({\cal G}_{\alpha \beta }+{\cal G}_\alpha {\cal G}_\beta \right)\,,\nonumber\\
P_L\upsilon  & =-\frac{1}{\sqrt{2}}\chi ^\alpha  \rme^{{\cal G}/2} {\cal G} _\alpha \sqrt{\frac{W}{\overline{W}}}\,.\label{defnGquant}
\end{align}

The potential is split in a positive and a negative contribution
\begin{eqnarray}
  V&=&V_-+V_+\,,\nonumber\\
 V_-&=& -3\rme^{ {\cal K}}W \overline{W}=-3|m_{3/2}|^2
\,,\qquad  V_+\,=\,
\rme^{ {\cal K}}\nabla _\alpha  W
    g^{\alpha \bar \beta
}\overline\nabla _{\bar \beta }\overline{W}
\,.\label{potN1sugrafinal}
\end{eqnarray}

It will be useful to define also the dimensionless relation between the two parts of the potential
\begin{equation}
  X\equiv  \frac{V_+}{|m_{3/2}|^2}= \frac{\nabla _\alpha  W \overline\nabla^{\alpha }\overline{W}}{W\overline{W}}= 3+\frac{V}{|m_{3/2}|^2}\,.
 \label{defX}
\end{equation}
In the ${\cal G}$-formulation, this is
\begin{equation}
  X = {\cal G}_\alpha \bar {\cal G}^\alpha = {\cal G}_\alpha g^{\alpha \bar \beta } \bar {\cal G}_{\bar \beta }\,.
 \label{defXG}
\end{equation}
This allows to write the potential as
\begin{equation}
  V = \rme^{{\cal G}}(X-3)\,.
 \label{VGX}
\end{equation}

The derivative of the potential can be written as
\begin{align}
 V_\alpha \equiv  \partial _\alpha V &=\rme^{{\cal K}/2}\left(  -2\overline{m}_{3/2}\nabla
  _\alpha W+m_{\alpha \beta }\overline\nabla ^\beta \overline{W}\right) \nonumber\\
  &= \rme^{{\cal G}}\left[{\cal G}_{\alpha \beta }\overline{{\cal G}}^\beta +(X-2){\cal G}_\alpha \right] \,.
\label{dalphaV}
\end{align}

%%%%%%%%%%%%%%%%%%%%%%%%%%%%%%%%%%%%%%%%%%%%%%%%%%%%%%%%
%%\mciteSetMidEndSepPunct{;\space}{}{\relax}
%%%%%%%%%%%%%%%%%%%%%%%%%%%%%%%%%%%%%%%%%%%%%%%%%%%%%%%%
%\bibliography{supergravity}

\begin{thebibliography}{10}

\bibitem{Ferrara:1979wa}
S.~Ferrara, L.~Girardello  and F.~Palumbo, \emph{{A general mass formula in
  broken supersymmetry}}, Phys. Rev. {\bf D20} (1979)
\href{http://dx.doi.org/10.1103/PhysRevD.20.403}{403}
%%CITATION = PHRVA,D20,403;%%.

\bibitem{Cremmer:1982en}
E.~Cremmer, S.~Ferrara, L.~Girardello  and A.~Van~Proeyen, \emph{{Yang--Mills
  theories with local supersymmetry: Lagrangian, transformation laws and
  superhiggs effect}}, Nucl. Phys. {\bf B212} (1983)
\href{http://dx.doi.org/10.1016/0550-3213(83)90679-X}{413}
%%CITATION = NUPHA,B212,413;%%.

\bibitem{Grisaru:1982sr}
M.~T. Grisaru, M.~Ro\v{c}ek  and A.~Karlhede, \emph{{The superhiggs effect in
  superspace}}, Phys. Lett. {\bf B120} (1983)
\href{http://dx.doi.org/10.1016/0370-2693(83)90634-2}{110}
%%CITATION = PHLTA,B120,110;%%.

\bibitem{Freedman:2012zz}
D.~Z. Freedman and A.~Van~Proeyen, {\em Supergravity}.
\newblock Cambridge University Press,
2012.
\newblock
%%CITATION = INSPIRE-1123253;%%.

\bibitem{Kallosh:2000ve}
R.~Kallosh, L.~Kofman, A.~D. Linde  and A.~Van~Proeyen, \emph{Superconformal
  symmetry, supergravity and cosmology}, Class. Quant. Grav. {\bf 17} (2000)
  4269--4338, \href{http://arXiv.org/abs/hep-th/0006179}{{\tt hep-th/0006179}},
erratum \textbf{21} (2004) 5017
%%CITATION = HEP-TH 0006179;%%.

\bibitem{Ferrara:2013pla}
S.~Ferrara, A.~Kehagias  and M.~Porrati, \emph{{Vacuum structure in a chiral
  ${\cal R}+{\cal R}^n$ modification of pure supergravity}}, Phys. Lett. {\bf
  B727} (2013)
  \href{http://dx.doi.org/10.1016/j.physletb.2013.10.027}{314--318},
\href{http://arxiv.org/abs/1310.0399}{{\tt arXiv:1310.0399 [hep-th]}}
%%CITATION = ARXIV:1310.0399;%%.

\bibitem{Ferrara:2016vzg}
S.~Ferrara and D.~Roest, \emph{{General sGoldstino Inflation}},
\href{http://arxiv.org/abs/1608.03709}{{\tt arXiv:1608.03709 [hep-th]}}
%%CITATION = ARXIV:1608.03709;%%.

\bibitem{Komargodski:2009rz}
Z.~Komargodski and N.~Seiberg, \emph{{From linear SUSY to constrained
  superfields}}, JHEP {\bf 0909} (2009)
  \href{http://dx.doi.org/10.1088/1126-6708/2009/09/066}{066},
\href{http://arxiv.org/abs/0907.2441}{{\tt arXiv:0907.2441 [hep-th]}}
%%CITATION = ARXIV:0907.2441;%%.

\bibitem{Kallosh:2016hcm}
R.~Kallosh, A.~Karlsson, B.~Mosk  and D.~Murli, \emph{{Orthogonal nilpotent
  superfields from linear models}}, JHEP {\bf 05} (2016)
  \href{http://dx.doi.org/10.1007/JHEP05(2016)082}{082},
\href{http://arxiv.org/abs/1603.02661}{{\tt arXiv:1603.02661 [hep-th]}}
%%CITATION = ARXIV:1603.02661;%%.

\bibitem{Volkov:1973ix}
D.~Volkov and V.~Akulov, \emph{{Is the neutrino a Goldstone particle?}}, Phys.
  Lett. {\bf 46B} (1973)
\href{http://dx.doi.org/10.1016/0370-2693(73)90490-5}{109--110}
% .

\bibitem{Casalbuoni:1988xh}
R.~Casalbuoni, S.~De~Curtis, D.~Dominici, F.~Feruglio  and R.~Gatto,
  \emph{{Non-linear realization of supersymmetry algebra from supersymmetric
  constraint}}, Phys.Lett. {\bf B220} (1989)
\href{http://dx.doi.org/10.1016/0370-2693(89)90788-0}{569}
%%CITATION = PHLTA,B220,569;%%.

\bibitem{Bergshoeff:2015tra}
E.~A. Bergshoeff, D.~Z. Freedman, R.~Kallosh  and A.~Van~Proeyen, \emph{{Pure
  de Sitter supergravity}}, Phys. Rev. {\bf D92} (2015), no.~8,
  \href{http://dx.doi.org/10.1103/PhysRevD.93.069901,
  10.1103/PhysRevD.92.085040}{085040},
  \href{http://arxiv.org/abs/1507.08264}{{\tt arXiv:1507.08264 [hep-th]}},
[Erratum: Phys. Rev.D93,no.6,069901(2016)]
%%CITATION = ARXIV:1507.08264;%%.

\bibitem{Antoniadis:2014oya}
I.~Antoniadis, E.~Dudas, S.~Ferrara  and A.~Sagnotti, \emph{{The
  Volkov-Akulov-Starobinsky supergravity}}, Phys.Lett. {\bf B733} (2014)
  \href{http://dx.doi.org/10.1016/j.physletb.2014.04.015}{32--35},
\href{http://arxiv.org/abs/1403.3269}{{\tt arXiv:1403.3269 [hep-th]}}
%%CITATION = ARXIV:1403.3269;%%.

\bibitem{Ferrara:2014kva}
S.~Ferrara, R.~Kallosh  and A.~Linde, \emph{{Cosmology with nilpotent
  superfields}}, JHEP {\bf 1410} (2014)
  \href{http://dx.doi.org/10.1007/JHEP10(2014)143}{143},
\href{http://arxiv.org/abs/1408.4096}{{\tt arXiv:1408.4096 [hep-th]}}
%%CITATION = ARXIV:1408.4096;%%.

\bibitem{Kallosh:2014via}
R.~Kallosh and A.~Linde, \emph{{Inflation and uplifting with nilpotent
  superfields}}, JCAP {\bf 1501} (2015)
  \href{http://dx.doi.org/10.1088/1475-7516/2015/01/025}{025},
\href{http://arxiv.org/abs/1408.5950}{{\tt arXiv:1408.5950 [hep-th]}}
%%CITATION = ARXIV:1408.5950;%%.

\bibitem{Dall'Agata:2014oka}
G.~Dall'Agata and F.~Zwirner, \emph{{On sgoldstino-less supergravity models of
  inflation}}, JHEP {\bf 12} (2014)
  \href{http://dx.doi.org/10.1007/JHEP12(2014)172}{172},
\href{http://arxiv.org/abs/1411.2605}{{\tt arXiv:1411.2605 [hep-th]}}
%%CITATION = ARXIV:1411.2605;%%.

\bibitem{Ferrara:2016ajl}
S.~Ferrara, A.~Kehagias  and A.~Sagnotti,
  \href{http://dx.doi.org/10.1142/S0217751X16300441}{\emph{{Cosmology and
  supergravity}},} in {\em {Memorial meeting for Nobel laureate Professor Abdus
  Salam's 90th birthday Singapore, January 25-28, 2016}}, vol.~A31, p.~1630044.
\newblock 2016.
\newblock
\href{http://arxiv.org/abs/1605.04791}{{\tt arXiv:1605.04791 [hep-th]}}.
\newblock
% .

\bibitem{Kachru:2003aw}
S.~Kachru, R.~Kallosh, A.~Linde  and S.~P. Trivedi, \emph{De Sitter vacua in
  string theory}, Phys. Rev. {\bf D68} (2003) 046005,
\href{http://arxiv.org/abs/hep-th/0301240}{{\tt hep-th/0301240}}
%%CITATION = HEP-TH 0301240;%%.

\end{thebibliography}
%%%Included for WinEdt Gather Purpose (do not remove the comment line below:
%%             %input "C:\localtexmf\bibtex\bib\*.bib"
%%             %input "C:\Program Files\MiKTeX\texmf\bibtex\bib\*.bib"
%%             %input "C:\localtexmf\Bibtex\bib\*.bib"
%%\bibliographystyle{toinemcite}
%\bibliographystyle{toine}
\providecommand{\href}[2]{#2}\begingroup\raggedright\endgroup

\end{document}